\documentclass[%
 reprint,
superscriptaddress,
 amsmath,amssymb,
 aps,
]{revtex4-2}

\usepackage{graphicx}
\usepackage{dcolumn}
\usepackage{bm}

\begin{document}

\preprint{APS/123-QED}

\title{Multistable nematic liquid crystal pores \\ as reconfigurable optical vortex metasurfaces}

\author{Claire Dor{\'e}}
\affiliation{
Department of Chemical and Biomolecular Engineering,
University of Pennsylvania, Philadelphia, PA 19104, USA
}
\author{Brian Edwards}%
\affiliation{Department of Electrical and Systems Engineering, University of Pennsylvania, Philadelphia, Pennsylvania 19104, USA
}%

\author{Nader Engheta}%
\affiliation{Department of Electrical and Systems Engineering, University of Pennsylvania, Philadelphia, Pennsylvania 19104, USA
}%

\author{Kathleen J. Stebe}
\email{kstebe@engineering.upenn.edu}
\affiliation{
Department of Chemical and Biomolecular Engineering,
University of Pennsylvania, Philadelphia, PA 19104, USA
}

\date{\today}

\begin{abstract}
Nematic liquid crystals confined in cylindrical pores adopt multiple stable director topologies --- radial, escaped, and arches --- each imprinting a distinct optical signature on transmitted circularly polarized light, with topological charge set by the projected director winding number, $\ell = 2m$. Using Landau--de Gennes and Jones-calculus modeling validated against micron-scale experiments, we show that this mechanism persists as the pore diameter shrinks toward the optical wavelength. However, full-wave Maxwell simulations of periodic sub-wavelength pore arrays reveal that the resulting vortex is a near-field phenomenon only: each topology imprints a distinguishable phase-singular signature --- differing in core multiplicity, core width, and ring amplitude --- directly below the pore array, but this signature does not survive propagation into the far field, where diffraction and inter-pore interference erase it within about one wavelength. Landau--de Gennes simulations further demonstrate that an applied electric field switches the pore between topologies, and hence between their near-field optical signatures. Because each topology is locally stable, sub-wavelength NLC pore arrays constitute an electrically reconfigurable, non-volatile platform for topology-encoded near-field optical readout, with potential application to optical labeling and anti-counterfeiting.
\end{abstract}

\maketitle


\section{\label{sec:level1}Introduction}

Optical vortex beams, which carry orbital angular momentum in their helical phase fronts, are a versatile tool in modern optics, enabling diverse applications including mode-division multiplexing in high-capacity communications \cite{YanHMCW, BozinovicTOAM}, in situ particle trapping and rotation in optical tweezers \cite{FrieseOATT}, resolution beyond the diffraction limit in microscopy \cite{ShenOV3Y}, and the encoding of quantum states for quantum information. These beams are typically generated using either fixed, passive elements including spiral phase plates \cite{SuedaLBGW}, forked diffraction gratings \cite{HeckenbergGOOP}, and geometric-phase (Pancharatnam-Berry) elements such as q-plates \cite{MarrucciOSAM}, each of which imprints a topological charge set at fabrication, or reconfigurable approaches like spatial light modulators, which can switch between arbitrary phase patterns on demand \cite{ForbesCADO}.

Nematic liquid crystal (NLC) films provide another important route to generating optical vortex beams, exploiting umbilical defects that form above a threshold applied voltage. These defects impose an azimuthal director winding on transmitted light, converting circularly polarized light into a vortex beam of fixed topological charge with tunable efficiency and wavelength \cite{BrasseletECTD}, or, under linear polarization, generating a discrete array of concentric vortex quadrupoles independently tunable in radial position and orientation \cite{BrasseletTOVA}. Other topological defects, including disclination lines and skyrmions \cite{LoussertMTOA,TangHEDL}, have also been used to pattern phase fields. Nematic droplets have also been exploited; a radial point defect confined within a micron-sized droplet imprints the same spin-to-orbital conversion mechanism, and has been demonstrated for both monochromatic and polychromatic light \cite{BrasseletOVFL}.  

NLCs offer special opportunities to generate, address, and reconfigure self-assembled defect arrays in confined settings. The combination of nonlinear elasticity, surface anchoring, and confinement gives NLCs a rich nonconvex energy landscape, enabling multistable director topologies; the same confinement can support multiple metastable, topologically distinct director configurations. Furthermore, their fluidity and responsiveness to external electric fields, flow, and temperature make them highly reconfigurable, allowing field-driven states to emerge or allowing the domain to switch between stable states \cite{BarbozaOVIV, KaraMPTI,KimMOTD,AplincPNMF,SasakiLSOR}. NLCs are also strongly birefringent, yielding significant retardance in sub-micron thicknesses while remaining optically transparent.

Thus far, research has targeted NLC structures with length scales of several microns, much greater than the wavelength of visible light (400–700 nm), so that the structure can be treated as a locally uniform, slowly varying optical element, well described by Jones/extended-Jones calculus. This raises the question of whether shrinking this length scale toward or below the optical wavelength could open new opportunities for encoding information within NLC structures. Conveniently, this sub-wavelength regime is also where two rigorous simulation methods become jointly tractable: Landau-de Gennes theory, which self-consistently resolves singular defect cores, is limited by mesh requirements near the core to domains of order 1 $\mu$m \cite{RavnikLGMO,BajcAMAS}, while full-wave electromagnetic methods (FDTD/FEM), which impose no paraxial assumption, are limited by 3D sub-wavelength anisotropy-gradient meshing to a few wavelengths \cite{HwangCMOT}. Around 500 nm–1 $\mu$m, then, both methods can be run on the same structure, making this scale the natural place to obtain quantitative, cross-validated predictions of optical response. Yet it remains largely unexplored: experimental and applied work has instead concentrated on structures above ~10 $\mu$m, the regime where Jones/extended-Jones calculus — a local, paraxial effective description valid for slowly-varying director fields — is trusted \cite{LienEJMR, GuEJMM}. Full-wave electromagnetic simulations of light propagation through nematic defects have been carried out \cite{HwangCMOT}, but not for the circularly polarized, vortex-generating configurations considered here, nor connected to the large-scale paraxial regime actually used experimentally; where the paraxial approximation breaks down as such structures shrink toward the optical wavelength therefore remains open.

Here, we study three distinct topological structures formed by NLCs confined in a periodic array of sub-wavelength circular pores with perpendicular (homeotropic) anchoring, and map the ability to reconfigure between these states using external fields. We then extend this analysis from the director structure to its optical consequences, comparing Jones-calculus predictions against full electromagnetic simulation of the same sub-wavelength pores to characterize where and why the two methods disagree. At this length scale, as expected, no far-field vortex beam is generated; instead, we find that (i) the full-wave near-field core is systematically broader than predicted by Jones calculus, with the surrounding bright ring extending into the nominally isotropic photoresist --- a direct signature of diffraction that the local, paraxial Jones model cannot capture; (ii) this near-field signature does not survive propagation into the substrate, where interference between the diffracted orders of neighboring pores in the periodic array replaces it with a multi-lobed, alternating-phase pattern within about one wavelength; and (iii) despite this loss of far-field character, each topology retains a distinguishable near-field optical signature --- differing in core multiplicity, core width, and ring amplitude --- directly below the pore array, and this signature switches electrically between topologies. Because each topology is a locally stable, non-volatile state, this combination of electrical writability and near-field optical readability suggests applications such as reconfigurable, non-volatile optical labeling, e.g.\ bar-coding materials to guard against counterfeiting.

\section{\label{sec:MaterialsandMethods}Materials \& Methods}

We first compute equilibrium and field-driven NLC configurations using Landau–de Gennes simulations. These Q-tensor fields are then used in two optical models. A pixel-wise Jones-calculus model provides an intuitive thin-element description of polarization conversion, while full-wave finite-element simulations solve Maxwell’s equations in the complete nanostructured geometry to assess diffraction, near-field structure, and substrate effects.
 
\subsection{Landau--de Gennes simulations}
 
The director structure is modeled within Landau--de Gennes theory, which describes nematic order via a tensor order parameter and can capture topological defects \cite{RavnikLGMO}; we summarize the main elements below. Nematic order is described by the $\mathbf{Q}$-tensor, a symmetric traceless $3\times3$ matrix related to the second moment of the molecular orientation distribution. The director field $\mathbf{n}$ and scalar order parameter $-1/2<S<1$ are obtained from the eigenvalue of $\mathbf{Q}$ with the largest absolute value, $\lambda_1=\frac{2}{3}S$, and its associated unit eigenvector $\mathbf{e}_1=\mathbf{n}$. In uniaxial regions, $Q_{ij}=S(n_in_j-\frac{1}{3}\delta_{ij})$.
 
The total free energy,
\begin{equation}
F_\text{tot}[\mathbf{Q}]=\int_\Omega f_\text{bulk}\,\mathrm{d}V+\int_{\partial\Omega}f_\text{surf}(\mathbf{Q})\,\mathrm{d}S,
\end{equation}
comprises a bulk contribution and a surface contribution that enforces the anchoring boundary condition. The bulk free energy density sums a phase (short-range order), elastic (distortion), and dielectric term:
\begin{equation}
f_\text{bulk}=f_\text{phase}+f_\text{elastic}+f_\text{dielectric}.
\end{equation}
 
Defined relative to the isotropic phase ($\mathbf{Q}=0$),
\begin{equation}
f_\text{phase}=\frac{A}{2}\mathrm{Tr}(\mathbf{Q}^2)+\frac{B}{3}\mathrm{Tr}(\mathbf{Q}^3)+\frac{C}{4}\big(\mathrm{Tr}(\mathbf{Q}^2)\big)^2.
\end{equation}
For 5CB in the nematic phase we use $A=-1.72\times10^5$, $B=-2.12\times10^6$, $C=1.73\times10^6~\mathrm{J/m^3}$, giving a ground-state order parameter $S_0=0.8$. The free-energy excess of the isotropic phase relative to the nematic ground state, $f_0=f_\text{phase}(0)-f_\text{phase}(S_0)$, is $3.84 \times 10^4 ~ \mathrm{J.m^{-3}}$.
 
For simplicity, we use the one-constant approximation,
\begin{equation}
f_\text{elastic}=\frac{L_1}{2}\partial_kQ_{ij}\partial_kQ_{ij}, \qquad L_1=10~\mathrm{pN},
\label{eq:elastic}
\end{equation}
which assigns equal cost to splay, twist, bend, and saddle-splay. The nematic coherence length is $\xi_\text{N}=\sqrt{L_1/f_0}$ and is an intrinsic length scale that sets the mesh resolution required to resolve defect cores.
 
In an electric field $\mathbf{E}$, a volume element of NLC acquires an induced polarization density $\mathbf{P}_\text{I}$ that depends on $\mathbf{Q}$ and $\mathbf{E}$; the dielectric free energy density is the corresponding electrostatic energy,
\begin{equation}
f_\text{dielectric}=-\frac{1}{2}\epsilon_0(\mathbf{\epsilon}_r\mathbf{E})\cdot\mathbf{E},
\label{eq:dielectric}
\end{equation}
where the relative permittivity tensor is related to molecular orientation by
\begin{equation}
\mathbf{\epsilon}_r=\epsilon_a\mathbf{Q}+\epsilon_\text{iso}\mathbf{I}, \qquad \epsilon_a=\frac{\epsilon_\parallel-\epsilon_\perp}{S_0},\quad \epsilon_\text{iso}=\tfrac{1}{3}(\epsilon_\parallel+2\epsilon_\perp).
\end{equation}
We assume only the DC field affects the director; the visible-wavelength EM wave propagating through the NLC is too weak to reorient it. For the DC field, $\epsilon_{r,\perp,\text{DC}}=6$ and $\epsilon_{r,\parallel,\text{DC}}=18$.
 
\paragraph{Surface free energy.} Homeotropic anchoring is enforced via the Rapini--Papoular potential,
\begin{equation}
f_\text{surf}=\frac{W}{2}\mathrm{Tr}(\mathbf{Q}-\mathbf{Q}_\text{S})^2,
\end{equation}
where $\mathbf{Q}_{\text{S},ij}=S_\text{S}(p_ip_j-\frac{1}{3}\delta_{ij})$ specifies the preferred director $\mathbf{p}$ and order parameter $S_\text{S}=0.9$ at the boundary. We use $W=10^{-2}~\mathrm{J/m^2}$, corresponding to strong anchoring at the sub-micron length scales simulated here.
 
\paragraph{Relaxation dynamics.} Each independent component of $\mathbf{Q}$ evolves by gradient descent,
\begin{equation}
\begin{split}
\gamma\,\partial_tQ_{ij} &= -\left[\frac{\partial f_\text{bulk}}{\partial Q_{ij}}-\nabla\cdot\frac{\partial f_\text{bulk}}{\partial\nabla Q_{ij}}\right] \quad \forall\,\mathbf{r}\in\Omega,\\
\gamma_\text{S}\,\partial_tQ_{ij} &= -\left[\frac{\partial f_\text{bulk}}{\partial\nabla Q_{ij}}\cdot\mathbf{\nu}+\frac{\partial f_\text{surf}}{\partial Q_{ij}}\right] \quad \forall\,\mathbf{r}\in\partial\Omega,
\end{split}
\label{eq:GLR}
\end{equation}
where $\mathbf{\nu}$ is the outward unit normal to $\partial\Omega$, and $\gamma$, $\gamma_\text{S}$ are the bulk and surface rotational viscosities. We use $\gamma=0.087~\mathrm{Pa\cdot s}$ and $\gamma_\text{S}=0.087\times10^{-6}~\mathrm{Pa\cdot m\cdot s}$, chosen so the boundary response time $\tau_\text{boundary}=\gamma_\text{S}/W$ is shorter than the bulk elastic response time $\tau_\text{elastic}=\gamma\ell^2/L_1$ at the length scale $\ell\approx250$~nm considered here. The radial, escaped, and arches configurations were obtained by initializing the relaxation with appropriate analytical director-field ansatzes.

\paragraph{Numerical implementation.} The governing equations are solved in dimensional form in COMSOL Multiphysics (Mathematics Module, General Form PDE, finite-element method), using a time-dependent study to simulate the gradient descent.
 
The simulation box has dimensions $L_x=L_y=1~\mu\mathrm{m}$, $L_z=500~\mathrm{nm}$, with periodic boundary conditions at $x=\pm L_x/2$ and $y=\pm L_y/2$. A photoresist layer of thickness $d=250~\mathrm{nm}$ contains a circular pore of diameter $D=500~\mathrm{nm}$. Strong homeotropic anchoring ($W=10^{-2}~\mathrm{J/m^2}$) is applied at the top surface ($z=L_z$), the photoresist surface, and the bottom of the pore.
 
The mesh is triangular on boundaries and tetrahedral in the bulk. The system is relaxed in two stages of 1~ms each: first on a uniform mesh, then on a mesh refined near the defect, with local element size $\max\!\big(\tfrac{1}{5}\sqrt{L_1/f_\text{elastic}},\,\xi_\text{N}\big)$. Mesh-refinement tests confirmed the defect structure is insensitive to further refinement.

The $\mathbf{Q}$ tensor components in the final state are exported on a regular grid, so that custom python code carries Jones calculus through the NLC slab.
 
\subsection{Jones calculus simulations}

We computed the optical response of the nematic liquid crystal (NLC) configurations using a pixel-wise Jones-calculus propagation model \cite{EllisSOPM}. We use the complex-field convention
\begin{equation}
\mathbf{E}(\mathbf{r},t)=\mathrm{Re}\left[\mathbf{E}_0 e^{i(kz-\omega t)}\right]
\end{equation}
for a plane wave propagating along $+z$, where $\mathbf{E}_0$ is the Jones vector, representing the polarization of the wave in the Cartesian polarization basis $\mathbf{e_x},\mathbf{e_y}$. With this convention, for propagation along $+z$, right circular polarization (RCP) and left circular polarization (LCP) are defined as

\begin{equation}
\mathbf{E}_{\mathrm{RCP}}=\frac{1}{\sqrt{2}}
\begin{pmatrix}
1\\
-i
\end{pmatrix},
\qquad
\mathbf{E}_{\mathrm{LCP}}=\frac{1}{\sqrt{2}}
\begin{pmatrix}
1\\
i
\end{pmatrix}.
\end{equation}
Thus, when viewed by a receiver looking back toward the source, the real electric field of the RCP state rotates clockwise, while that of the LCP rotates counterclockwise. 

The NLC is treated as a locally uniaxial birefringent material with uniform scalar order parameter. At each voxel, the local optical axis $\mathbf{n}(x,y,z)$ is obtained by diagonalizing the simulated $\mathbf{Q}$-tensor and selecting the eigenvector associated with the largest eigenvalue. Note that the Jones-calculus model uses only the director orientation and neglects the scalar-order reduction in defect cores. The NLC slab is discretized into $N_z=50$ layers of thickness $\Delta z=10 \ \mathrm{nm}$, and into $N_x=100$ and $N_y=100$ pixels in the transverse directions. Each voxel is modeled as a local linear retarder whose slow axis is the projection of $\mathbf{n}$ onto the transverse $x,y$ plane. If $\theta$ is the angle between $\mathbf{n}$ and the propagation axis $z$, the effective extraordinary refractive index for normally incident light is
\begin{equation}
n_{\mathrm{eff}}(\theta)=\frac{n_e n_o}
{\sqrt{n_e^2\cos^2\theta+n_o^2\sin^2\theta}},
\label{Eq:n_eff}
\end{equation}
where $n_o$ and $n_e$ are the ordinary and extraordinary refractive indices. The retardance accumulated over one voxel is therefore
\begin{equation}
    \delta(x,y,z;\lambda)=\frac{2\pi}{\lambda}
\left[n_{\mathrm{eff}}(\theta)-n_o\right]\Delta z .
\label{Eq:retardation}
\end{equation}

Writing $\phi=\mathrm{atan2}(n_y,n_x)$ for the azimuthal angle of the projected director, the Jones matrix of one voxel, expressed in the fixed Cartesian basis, is

\begin{equation}
\mathbf{J}(x,y,z;\lambda)=
\mathbf{R}^{-1}(\phi)
\begin{pmatrix}
e^{i\delta(x,y,z;\lambda)} & 0\\
0 & 1
\end{pmatrix}
\mathbf{R}(\phi),
\label{Eq:JonesMat}
\end{equation}

with

\begin{equation}\mathbf{R}(\phi)=
\begin{pmatrix}
\cos\phi & \sin\phi \\
-\sin\phi & \cos\phi
\end{pmatrix}.
\end{equation}

Regions outside the NLC are assigned the identity Jones matrix. For each transverse pixel $x,y$, the total Jones matrix of the stack is obtained by multiplying the voxel matrices in the order encountered by the propagating beam, and the transmitted field is

\begin{equation}
\mathbf{E}_{\mathrm{out}}(x,y;\lambda)
=
\mathbf{J}_{N_z-1}(x,y;\lambda)
\cdots
\mathbf{J}_{1}(x,y;\lambda)
\mathbf{J}_{0}(x,y;\lambda)
\mathbf{E}_{\mathrm{in}}.
\end{equation}

Analyzers are applied by projecting $\mathbf{E}_{\mathrm{out}}$ onto a chosen polarization state $\mathbf{a}$, giving the detected complex amplitude

\begin{equation}
E_{\mathrm{out}, a}(x,y;\lambda)=
\mathbf{a}^{\dagger}\mathbf{E}_{\mathrm{out}}(x,y;\lambda),
\end{equation}

and the corresponding amplitude
\begin{equation}
    A_a(x,y;\lambda)=|\mathbf{E}_{\mathrm{out}, a}(x,y;\lambda)|,
\end{equation}

and phase

\begin{equation}
    \Phi_a(x,y;\lambda)=\mathrm{arg}[\mathbf{E}_{\mathrm{out}, a}(x,y;\lambda)].
\end{equation}

In the simulations reported in Fig. \ref{figure:fig2}, light propagates from the top of the structure toward the bottom, i.e. along $-z$, so that it first crosses the NLC layer and then the photoresist layer. This is implemented by applying the Jones matrices in the corresponding order along the optical path. We used a uniform incident LCP beam and analyzed the RCP component of the output field. We used   $n_o=1.5$ for the ordinary index and $n_e=1.7$ for the extraordinary index, matching the values for the 5CB nematic phase.

For white-light simulations, shown in Fig. \ref{figure:fig1} to be compared with experimental images, the propagation is repeated independently for a set of wavelengths spanning the visible spectrum, from 380 to 780 nm with 5 nm spacing \cite{ZouICON, ChenLCPO}. The wavelength-dependent transmitted intensity is weighted by the input illumination spectrum and converted to CIE XYZ tristimulus values using the CIE 1931 color-matching functions. The resulting XYZ image is then transformed to RGB for visualization.

\subsection{Full-wave simulations}
 
Full-wave solutions of Maxwell's equations are obtained in the frequency domain using the Electromagnetic Waves module of COMSOL Multiphysics. The complex electric field satisfies
\begin{equation}
\nabla\times\mu_r^{-1}(\nabla\times\mathbf{E})-k_0^2\mathbf{\epsilon}_{r,\lambda}\mathbf{E}=0,
\end{equation}
with $\mu_r=1$ everywhere. In the NLC, the relative permittivity is the tensor field $\mathbf{\epsilon}_{r,\lambda}=\epsilon_a\mathbf{Q}+\epsilon_\text{iso}\mathbf{I}$ (as defined above), with $\epsilon_{r,\perp,\lambda}=2.4$ and $\epsilon_{r,\parallel,\lambda}=3.1$, which set the scale of the optical anisotropy. Because the $\mathbf{Q}$ tensor is used directly, the effective birefringence is weighted by the simulated scalar order parameter $S$, and is reduced in the defect core. All other domains (substrates, photoresist, PML) are isotropic with $\epsilon_r=2.4$. The refractive indices follow from $n=\sqrt{\epsilon_r}$.
 
The system has a layered geometry in $z$: the LC cell (photoresist + NLC) is sandwiched between two glass layers of thickness $L_\text{sub,sup}=\lambda_0/n_e$, themselves bounded by two perfectly matched layers (PML) of thickness $\lambda_0/2$ to suppress reflections, with periodic boundary conditions in $x$ and $y$. The wave propagates along $-z$, and the input port is defined at the interface between the upper PML and the superstrate.
 
The mesh is built sequentially: a swept square mesh in the PMLs, followed by triangular meshes on all interfaces. The NLC bulk is meshed with tetrahedral, refined near defects to $\min\!\big(\max(\tfrac{1}{5}\sqrt{L_1/f_\text{elastic}},\,\xi_\text{N}),\,\lambda_0/10\big)$. Everywhere, the mesh size $\ell_\text{EM}$ is bounded by $\lambda_0/100<\ell_\text{EM}<\lambda_0/10$.
 
We use the interface between the upper PML and the superstrate as the input plane, where a port launches an LCP wave propagating along $-z$. After solving the frequency-domain problem, the total complex electric field is exported on a regular three-dimensional grid and post-processed in Python. The transverse field is projected onto the RCP basis state to extract the phase and amplitude of the converted circularly polarized component. For convenience, we define the axial coordinate $\zeta=L_z-z$, so that increasing $\zeta$ follows the propagation direction. The RCP component is plotted on two transverse planes, as indicated in Fig.~\ref{figure:fig3}A.

\subsection{Experiments}
Circular wells were fabricated in 5~$\mu$m thick SU-8 films patterned on glass substrates by standard photolithography. The wells had a diameter of 15~$\mu$m. After fabrication, the substrates were treated with dimethyl octadecyl[3-(trimethoxysilyl)propyl] ammonium chloride (DMOAP) to promote homeotropic anchoring. Cells were assembled using a DMOAP-treated glass lid and 50~$\mu$m spacers, then filled with 5CB in the isotropic phase and quenched to the nematic phase. The resulting director configurations were imaged using a Zeiss Axio Vert A1 inverted microscope under white-light illumination, both in bright field and between crossed polarizers.

\section{\label{sec:Results}Results}

\begin{figure*}
\includegraphics{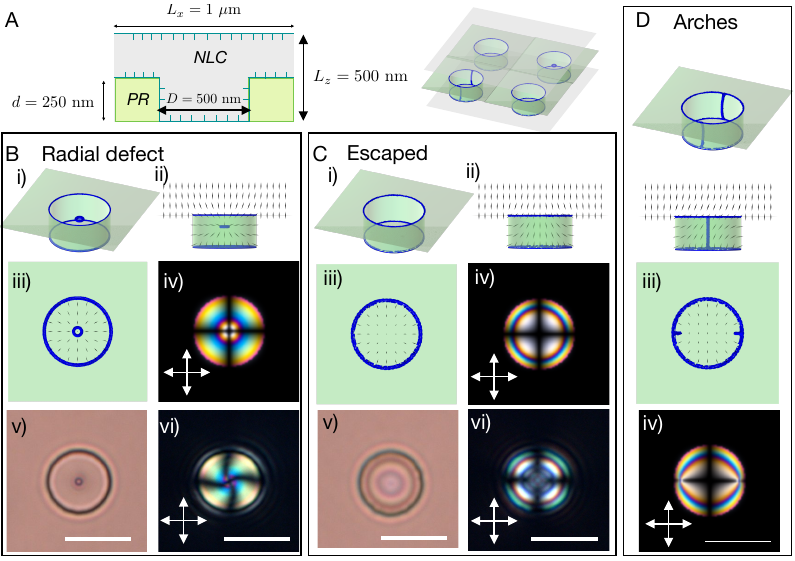}
\caption{\textbf{Multistable director topologies in nematic liquid crystal pores.} (A) Schematic of the platform: nematic liquid crystal (NLC) confined in circular photoresist (PR) pores, with homeotropic (perpendicular) anchoring at the top and bottom substrates and lateral confinement imposed by the pore wall. (B-D) Three representative metastable topologies: (B) radial defect, (C) escaped structure, (D) arches structure. Each state is shown through 3D director-field renderings (i), vertical cross-sections (ii), top-view director maps (iii), simulated polarized optical textures under white light illumination (iv), and for the radial defect and the escaped state, experimental microscopy images under Bright Field (v) and crossed-polarized microscopy (vi). Singular defect regions, identified by $S<0.7$, are shown in blue. For v) and vi), scale bar = 15 um. }
\label{figure:fig1}
\end{figure*}

\subsection{Multistable director topologies in NLC pores}
 
Landau--de Gennes relaxation within the pore geometry described above yields three distinct metastable director configurations, depending on the initial condition (Fig.~\ref{figure:fig1}): a \textit{radial defect} state, in which a disclination ring of winding number $+\frac{1}{2}$ encircles the pore axis (Fig.~\ref{figure:fig1}B); an \textit{escaped} state, in which the director rotates continuously out of the pore plane without a singular core (Fig.~\ref{figure:fig1}C); and an \textit{arches} state, in which the ring is replaced by two disclination arcs connecting the top and bottom edges of the pore (Fig.~\ref{figure:fig1}D). Singular regions, identified by $S<0.7$, are confined to a narrow ring for the radial defect and to the two arcs for the arches state; no such region appears in the bulk for the escaped state that possesses a non-singular director field.
 
Top-view director maps (Fig.~\ref{figure:fig1}, panels iii) show that the radial defect and escaped states both retain the circular symmetry of the pore, while the arches state breaks this symmetry with its two defect arcs. Correspondingly, the simulated white-light polarized textures (panels iv) are close to concentric for the radial and escaped states but asymmetric for the arches state.
 
We recover in experiments of 5CB-filled circular wells (15~$\mu$m diameter, 5~$\mu$ m thick — roughly 30$\times$ larger than the simulated pores) two kinds of configurations compatible with the radial defect and the escaped states, imaged under bright-field and crossed-polarized illumination (Fig.~\ref{figure:fig1}B--C, panels v--vi). Their crossed-polarized textures in particular reproduces the concentric fringe pattern and interference colors seen in simulation (Fig.~\ref{figure:fig1}B--C, panels iv), confirming that these topologies are physically accessible at both the sub-wavelength and micron scales despite the difference in absolute size. In the experiments, the radial defect configuration exhibits a chiral texture, which we attribute to the development of twist near the defect core to reduce the large splay elastic distortion. Such spontaneous chiral symmetry breaking is expected in 5CB because twist deformations are energetically less costly than splay ($K_{22}<K_{11}$), consistent with previous theoretical and experimental studies \cite{LavrentovichPTAT,RudingerTTIN}. In contrast with the radial defect and escape states, the arches configuration is not observed in our micron-scale experimental pores.

\subsection{Optical vortex generation predicted by Jones calculus}

\begin{figure*}
\includegraphics{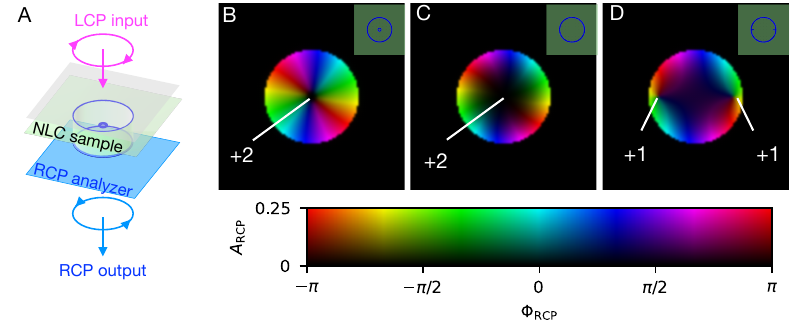}
\caption{Optical vortex beam generation predicted by Jones calculus.
(A) Schematic of the optical conversion: an incident left-circularly polarized (LCP) plane wave is transmitted through the NLC pore and analyzed in the right-circularly polarized (RCP) channel. (B-D) Phase-amplitude maps of the RCP output field for three director topologies in 250 nm-deep pores: (B) radial defect, (C) escaped radial state, and (D) arches. Hue encodes the optical phase and brightness encodes the field amplitude. The radial and escaped states both generate an RCP optical vortex beam with total topological charge +2, appearing as a bright annulus surrounding a dark central core. The radial defect produces a narrow vortex core, approximately 70 nm wide, whereas the escaped state produces a broader core. In the arches state, the output field separates into two +1 phase singularities located near the arch disclination lines. Calculations were performed at $\lambda=635$ nm.}
\label{figure:fig2}
\end{figure*}
 
Pixel-wise Jones calculus (Fig.~\ref{figure:fig2}) shows that an incident LCP plane wave acquires a nonzero RCP component — and thus orbital angular momentum — after passing through the NLC pore, for all three topologies. For the radial defect and escaped states, the RCP output field forms a single optical vortex of total topological charge $+2$: a bright annulus of near-uniform amplitude surrounds a dark, phase-singular core (Fig.~\ref{figure:fig2}B--C). Although these two states differ in three-dimensional director topology — the radial defect contains a singular disclination ring, whereas the escaped state removes this singularity by tilting the director out of the transverse plane along the pore axis — they yield the same optical charge because the Jones-calculus construction is sensitive only to the director's projection onto the transverse $(x,y)$ plane at each depth (Eq.~\ref{Eq:n_eff},\ref{Eq:retardation},\ref{Eq:JonesMat}); it is the winding number of this projected, effectively two-dimensional field, and not the full 3D topology of $\mathbf{n}(x,y,z)$, that sets the topological charge of the transmitted vortex. The two states differ mainly in the size of this core, approximately 70~nm for the radial defect (the diameter of the singular ring) versus a broader core for the escaped state, reflecting the concentrated director gradient at the disclination ring compared to the smoother, non-singular splay of the escaped configuration. In the arches state (Fig.~\ref{figure:fig2}D), the total charge-$+2$ core splits into two separated charge-$+1$ singularities located below the two arch disclination lines: the total topological charge is conserved, but its spatial distribution, and hence the transverse structure of the output beam, is qualitatively different.
 
These results, obtained at $\lambda = 635$~nm for $d=250$~nm-deep pores, indicate that all three multistable topologies convert circular polarization into orbital angular momentum, but encode it in distinct spatial mode structures (one charge-$+2$ vortex versus two charge-$+1$ vortices), providing a basis for optically discriminating between reconfigurable states.

\subsection{Full-wave validation at sub-wavelength scale}

\begin{figure*}
\includegraphics{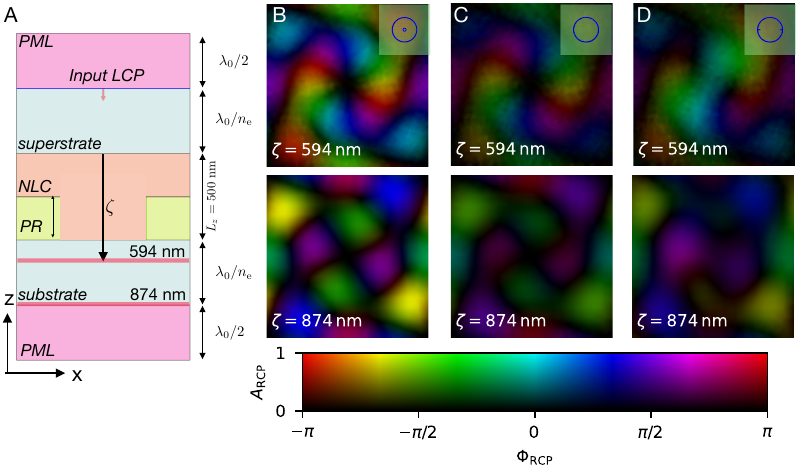}
\caption{\textbf{Full-wave Maxwell simulations validate near-field optical vortex at sub-wavelength scale} A) simulation set-up for the full-wave Maxwell simulation through the square lattice of circular pores. (B-D) Phase-amplitude maps of the RCP output field in two different focal planes at distance $\zeta$ from the upper interface of the NLC, for three director topologies in 250 nm-deep pores: (B) radial defect, (C) escape state and (D) arches state. Calculation were performed for $\lambda = 635$ nm.}
\label{figure:fig3}
\end{figure*}

The Jones-calculus model above is both paraxial and, more restrictively, \textit{local}: it is pixel-wise, so the field exiting at transverse position $(x_k,y_k)$ depends only on the sequence of local retardances and optical-axis orientations encountered directly above it, with no dependence on the director field at any neighboring column $(x_{k'},y_{k'})$. This is a stronger locality than that of Maxwell's equations themselves: the wave equation is a local partial differential equation, but because it involves spatial derivatives, the field at a point remains coupled to its infinitesimal neighborhood, and integrating that coupling over a finite propagation distance is precisely what produces diffraction — the exchange of amplitude and phase between transverse points. The validity of the Jones-calculus approximation therefore does not depend on the sample being ``thin'' in an absolute sense, but on the lateral length scale over which the director field varies being large compared to the sample thickness, so that light has no opportunity to diffract from one column into its neighbor before exiting. Here, the director rotates over a lateral scale of tens to a few hundred nanometers near each defect core, comparable to or smaller than the $\sim$500~nm thickness of the pore and surrounding layers, so this condition is not well satisfied.
 
We therefore repeated the calculation with full-wave Maxwell simulations of a periodic square lattice of pores (Fig.~\ref{figure:fig3}A), extracting the RCP-projected field on transverse planes at two distances $\zeta$ from the upper NLC interface: $\zeta = 594$~nm, just below the NLC layer, and $\zeta = 874$~nm, further down within the substrate.
 
Close to the sample ($\zeta = 594$~nm), the radial defect state (Fig.~\ref{figure:fig3}B) and the escape state (Fig.~\ref{figure:fig3}C) retain a single phase-singular core about which the RCP phase winds smoothly, confirming that a vortex-like near field forms even at this sub-wavelength scale, in qualitative agreement with the Jones-calculus prediction. This core is, however, markedly wider than its Jones-calculus counterpart for the radial state (Fig.~\ref{figure:fig2}B): the central amplitude dip broadens. For both the radial defect and escape states, the bright ring surrounding the core is no longer confined to the pore diameter, extending laterally into the surrounding photoresist — beyond the region where the pixel-wise Jones model would predict any polarization conversion at all, since the photoresist is treated as optically isotropic. The amplitude of the bright ring is higher for the radial defect than for the escape state. The arches state (Fig.~\ref{figure:fig3}D) departs more strongly from its Jones-calculus prediction at the same plane: rather than the two well-separated $+1$ cores expected from the symmetry breaking of this topology (Fig.~\ref{figure:fig2}D), the individual cores are barely discernible, already partly obscured at $\zeta = 594$~nm.

Deeper into the substrate ($\zeta = 874$~nm), the radial defect and escape states develop multi-lobed, alternating-phase structure, and any simple vortex signature is lost. We attribute this evolution with propagation distance to interference between the near field diffracted by each pore and the fields scattered by its periodic neighbors under the periodic boundary conditions used here: close to the exit face the field is still dominated by the local pore, whereas at larger $\zeta$ the diffracted orders from neighboring pores in the lattice increasingly overlap and interfere, breaking the single-pore vortex into the lobed pattern seen at $\zeta = 874$~nm. This picture suggests a testable prediction: at larger pore-to-pore spacing, where neighboring pores contribute negligible scattered field within the sampled plane, the full-wave near field should more closely approximate that of an isolated pore. Whether that isolated-pore field then converges toward the Jones-calculus core width, or remains broadened by sub-wavelength diffraction as already seen at $\zeta = 594$~nm, is not established by the present simulations.

\subsection{Field-driven topological reconfiguration}

\begin{figure*}
\includegraphics{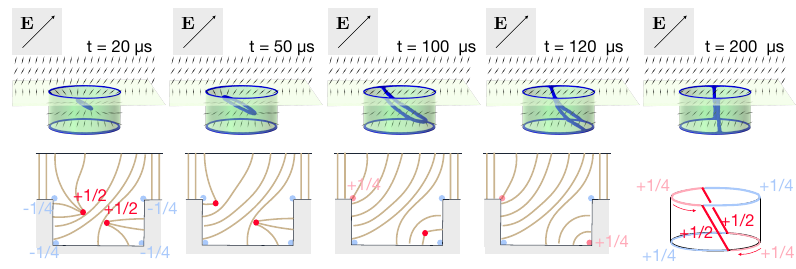}
\caption{Topological reconfiguration driven by an external electric field. Top row: director field in the $yz$ plane during the transition; bottom row: schematic cross sections of the pore in the same plane, with the winding number of each defect annotated. The pore is initially in the radial defect state, and a field of amplitude $E_0=5\times 10^6~\mathrm{V/m}$ is applied at $45^\circ$ obliquely in the $yz$ plane. The ring forming the radial defect state carries a uniform $+\frac{1}{2}$ winding number, while the pore edges carry a $-\frac{1}{4}$ geometric winding. Under the field (positive dielectric anisotropy, $\Delta \epsilon> 0$), the ring reorients and widens to maximize alignment with the field. The $+\frac{1}{2}$ ring partially recombines with the top and bottom edge disclination lines, locally changing the edge winding to $+\frac{1}{4}$; the unrecombined segments of the ring form arches connecting the two edges.}
\label{figure:fig4}
\end{figure*}
 
Finally, we tested whether an applied DC electric field can switch the pore between topologies. Starting from the radial defect state, we applied a uniform  and constant field of amplitude $E_0=5\times 10^6~\mathrm{V/m}$ oriented obliquely ($45^\circ$) in the $yz$ plane (Fig.~\ref{figure:fig4}). Because 5CB has positive dielectric anisotropy, the field drives the director toward alignment with $\mathbf{E}$. Over $\sim$200~\textmu s, the $+\tfrac{1}{2}$disclination ring reorients and widens to increase director alignment with the field. In doing so, it partially recombines with the disclination lines pinned at the top and bottom pore edges, locally converting their geometric winding number from $-\tfrac{1}{4}$ to $+\tfrac{1}{4}$; the segments of the ring that do not recombine persist as arcs connecting the two edges. By $t=200$~\textmu s, this process has converted the radial defect state into the arches state of Fig.~\ref{figure:fig1}D.
 
This field-driven transition demonstrates switching between two of the three multistable topologies and, by extension via Fig.~\ref{figure:fig2}, between their distinct optical vortex signatures (a single charge-$+2$ core versus two separated charge-$+1$ cores). The reverse transition (arches back to radial defect) is the object of future work.

\section{\label{sec:Discussion}Discussion}

Together, these results show that the multistability and field-driven reconfigurability established for NLC pores at the micron scale persist when the same topologies are confined to sub-wavelength dimensions, but the optical output changes character. Jones calculus and full-wave simulations agree that a vortex-like near field still forms directly below the pore array: the radial and escaped states each support a single phase-singular RCP core (Fig.~\ref{figure:fig2} and Fig.~\ref{figure:fig3}). Quantitatively, however, the two methods diverge in a specific and physically meaningful way. The full-wave core is broader than predicted by Jones model, and the surrounding bright ring extends into the photoresist, a region the pixel-wise Jones model treats as isotropic and therefore forbids from contributing to polarization conversion. This divergence is a direct signature of the diffraction that a strictly local, paraxial model cannot capture, consistent with our estimate that the director rotates over lateral scales comparable to the layer thickness itself. On propagation into the substrate, the vortex signature does not survive: interference between diffracted orders from neighboring pores in the periodic lattice replaces the single- or double-core structure with a multi-lobed, alternating-phase pattern within about one wavelength. No far-field vortex beam is therefore generated at this scale, in contrast to the micron-scale case.

This reframes the practical opportunity. Rather than a far-field vortex-beam source, sub-wavelength NLC pores offer topology-dependent near-field signatures — core multiplicity, core width, and ring amplitude — that are addressable through electric field-driven switching between metastable states, as demonstrated for the radial-to-arches transition (Fig.~\ref{figure:fig4}). Because each topology is a locally stable, non-volatile state, this combination of electrical writability and optical readability at sub-wavelength resolution is naturally suited to applications such as reconfigurable, non-volatile optical labeling. Two questions remain open: whether isolated pores at larger pitch recover a cleaner near-field vortex or remain broadened regardless of inter-pore spacing, and whether the reverse (arches-to-radial) and escaped-state transitions can also be field-driven. More broadly, this system offers a concrete example of where local, paraxial optical descriptions break down as patterned soft-matter structures shrink toward the optical wavelength.


\begin{acknowledgments}
This work was supported by the U.S. Army Research Office (ARO) under Grant No. W911NF2210205, ``Multistable Elastic Pixel (MEP)-Based Reconfigurable Electromagnetic Devices.''
\end{acknowledgments}

\bibliography{bibref}

\end{document}